
\magnification=1200
\def\ni{\noindent}
\def\.{\mathaccent 95}
\def\a{\alpha}
\def\be{\beta}
\def\ga{\gamma}

\def\ep{\epsilon}

\def\ka{\kappa}
\def\la{\lambda}

\def\si{\sigma}

\def\Si{\Sigma}

\def\Om{\Omega}

\def\frac#1#2{{\textstyle{{#1}\over {#2}}}}
\def\ni{\noindent}
\def\lsim{\mathrel{\rlap{\lower4pt\hbox{\hskip1pt$\sim$}}
    \raise1pt\hbox{$<$}}}
\def\gsim{\mathrel{\rlap{\lower4pt\hbox{\hskip1pt$\sim$}}
    \raise1pt\hbox{$>$}}}
\def\sqr#1#2{{\vcenter{\vbox{\hrule height.#2pt
         \hbox{\vrule width.#2pt height#1pt \kern#1pt
         \vrule width.#2pt}
         \hrule height.#2pt}}}}

\newbox\grsign \setbox\grsign=\hbox{$>$} \newdimen\grdimen \grdimen=\ht\grsign
\newbox\simlessbox \newbox\simgreatbox
\setbox\simgreatbox=\hbox{\raise.5ex\hbox{$>$}\llap
     {\lower.5ex\hbox{$\sim$}}}\ht1=\grdimen\dp1=0pt
\setbox\simlessbox=\hbox{\raise.5ex\hbox{$<$}\llap
     {\lower.5ex\hbox{$\sim$}}}\ht2=\grdimen\dp2=0pt

%
%

\def\ref#1  {\noindent \hangindent=24.0pt \hangafter=1 {#1} \par}
\def\doublespace {\smallskipamount=6pt plus2pt minus2pt
                  \medskipamount=12pt plus4pt minus4pt
                  \bigskipamount=24pt plus8pt minus8pt
                  \normalbaselineskip=24pt plus0pt minus0pt
                  \normallineskip=2pt
                  \normallineskiplimit=0pt
                  \jot=6pt
                  {\def\smallskip {\vskip\smallskipamount}}
                  {\def\medskip   {\vskip\medskipamount}}
                  {\def\bigskip   {\vskip\bigskipamount}}
                  {\setbox\strutbox=\hbox{\vrule 
                    height17.0pt depth7.0pt width 0pt}}
                  \parskip 12.0pt
                  \normalbaselines}
\def\ni{\noindent}
\def\ts{\times}

\def\bbfv{\bar{\bf V}}
\def\bbfb{\bar{\bf B}}
\def\bfbb{\bar{\bf B}}
\def\bfbv{\bar{\bf V}}
\def\ba{\bar A}
\def\bb{\bar B}
\def\sun{_{\odot}}
\def\ref{\par\noindent\hangindent 20pt}
\def\sles{\lower2pt\hbox{$\buildrel {\scriptstyle <} 
   \over {\scriptstyle\sim}$}}
\def\sgreat{\lower2pt\hbox{$\buildrel {\scriptstyle >} 
   \over {\scriptstyle\sim}$}}
\def\lapprox{\lower2pt\hbox{$\buildrel \lower2pt\hbox{${\scriptstyle<}$} 
   \over {\scriptstyle\approx}$}}
\def\gapprox{\lower2pt\hbox{$\buildrel \lower2pt\hbox{${\scriptstyle>}$} 
   \over {\scriptstyle\approx}$}}
\def\both{\lower2pt \hbox{$\buildrel {\leftarrow} \over {\rightarrow}$}}


\newbox\grsign \setbox\grsign=\hbox{$>$} \newdimen\grdimen \grdimen=\ht\grsign
\newbox\simlessbox \newbox\simgreatbox
\setbox\simgreatbox=\hbox{\raise.5ex\hbox{$>$}\llap
     {\lower.5ex\hbox{$\sim$}}}\ht1=\grdimen\dp1=0pt
\setbox\simlessbox=\hbox{\raise.5ex\hbox{$<$}\llap
     {\lower.5ex\hbox{$\sim$}}}\ht2=\grdimen\dp2=0pt
\def\simg{\mathrel{\copy\simgreatbox}}
\def\siml{\mathrel{\copy\simlessbox}}
\newbox\simppropto
\setbox\simppropto=\hbox{\raise.5ex\hbox{$\sim$}\llap
     {\lower.5ex\hbox{$\propto$}}}\ht2=\grdimen\dp2=0pt

\centerline{\bf CAN THE FORMATION OF X-RAY OBSCURING TORI AND JETS} 
\centerline{\bf IN ACTIVE GALAXIES BE DETERMINED BY ONE PARAMETER?}

\medskip\medskip

\centerline{\bf Eric G. Blackman}
\smallskip
\centerline{Institute of Astronomy, Madingley Road, Cambridge, CB3 OHA,
England}
\smallskip
\centerline{and}
\smallskip
\centerline{\bf Insu Yi}
\smallskip
\centerline{Institute for Advanced Study, Olden Lane, Princeton, NJ, 08540}
\bigskip
\centerline{(accepted for publication in {\it Astrophysical Journal Letters})}

\bigskip\bigskip
\centerline{\bf ABSTRACT}
\medskip\medskip
\doublespace

A torus of reduced differential rotation can form in the inner 
$\siml 10$pc core of active galactic nuclei incurring a density enhancement
that can account for obscuration of  X-rays in Seyferts  
when the initial inner core to black hole mass ratio  $\gsim 50$.
The same density enhancement and reduction in differential rotation 
can also lead to dynamo growth of 
poloidal fields which attain a magnitude $\sim 10^4$G when accreted onto 
the central engine.  As radio jet models often employ poloidal 
fields as agents in extracting power for the jet luminosity, we suggest 
that jetted AGN might require this poloidal field production.
Although the poloidal field would be originally
produced in the obscuring torus, jetted objects are less likely to have 
obscuring tori:  The poloidal field would only aid in powering jet emission 
after it accretes with the torus matter to the central engine.
Thus, only during the relatively short torus accretion time scale 
could there be both a jet and and torus.

\medskip
\noindent 
{\bf Subject Headings:} Galaxies: Active, Jets, Magnetic Fields, 
Radio Continuum: Galaxies
\vfill
\eject

\noindent
{\bf 1. Introduction}

At the centers of galaxies, and particularly 
in active galactic nuclei (AGN), 
are likely black holes (BH) which power 
the central luminosity by accretion
(e.g., Rees 1984).  Time variability and the 
estimated accretion efficiency seem to require black hole
masses  $\simg 10^7 M_{\sun}$ in many jetted and jet-free sources. 
Little is known about the detailed mass 
distribution in the ``inner core'' (IC) $\siml 10$pc 
regions  of all galaxy types, but there in our own
Galaxy the rotation curve seems to incur a change 
from Keplerian to flat (Genzel \& Townes 1987).  The transition region 
is quasi-rigidly rotating. This leads to a density enhancement over a purely 
Keplerian curve, as we show below, and may explain the presence of the 
cicumnuclear ring (CNR) 
of the Galaxy and X-ray obscuring Seyfert tori (Yi et al. 1994, Duschl 1989)
in accordance with unified models of AGN (e.g. Antonoucci, 1993).

We show that such a region also favors the dynamo production 
of poloidal magnetic field (PMF) to a magnitude which, when accreted onto an
AGN central engine, is likely
$\sim 10^4$G---in agreement with that 
inferred by other equipartition estimates 
(Begelman et al. 1984). Such PMF is often required in jet models 
(Blandford \& Znajek 1977, Lovelace et al. 1987, Appl \& Camenzind 1993, 
Lynden-Bell 1995).  The PMF dynamo growth time scale is much smaller than the  
torus depletion time, so whether significant PMF is 
produced depends only on the initial  to BH mass ratio. 
Though the jet PMF originates in the torus, the 
field can only play a role in jets after accreting to the central engine.
Only during the short time when the torus is depleting could there be 
both a jet and torus in one object.  This is consistent with the fact that
direct evidence for tori comes mainly from radio quiet 
objects (Urry and Padovani, 1995), but more data are needed.

\noindent
{\bf 2. Estimation of Time Scales and Adiabatic Black Hole Growth}

Rotation curves of spiral galaxies show quasi-rigid rotation 
in the inner $\sim$ kpc, and flat rotation curves outside $\sim$ kpc 
(Oort 1978, Binney \& Tremaine 1987).  
Models which account for observed  galaxy gas rotation curves 
seem to require (Binney \& Tremaine 1987)
(i) a central BH (ii) a stellar IC within a few-10pc, 
(iii) a more diffuse nuclear bulge of several 100pc, and
(iv) an isothermal sphere of dark matter on kpc scales.
Here we are interested in (i), the IC region sub-structure to 
the overall rotation curves, which is likely similar for
all galaxy types.

We first show that a typical BH grows by accretion slowly compared 
to orbital time scales, but rapidly compared to the 
IC relaxation time, so the hole's growth is nearly adiabatic:  
At any time, the accreting region will be in an approximately 
steady state if the 
viscous time scale $\tau_{vis}\ll \tau_{g}$, where $\tau_{g}$ is the 
BH growth time scale. An estimate for $\tau_{vis}$ is 
$$
\tau_{vis} \sim {R_d^2\over \nu} \sim {R_d\over V_r}
\sim {R_d\over (10^{-2}V_\phi)} 
\sim 5\times10^6 {\rm yr} \left(R_d\over 3\ts 10^{19}{\rm cm}\right)^{3/2}
\left(M_H\over 10^7M_\odot\right)^{-1/2},
\eqno(1)
$$
where $V_\phi$ is the rotational velocity, $V_r$ is the radial
velocity, and $R_d$ is radial extent of the accretion flow. We estimate 
the  BH growth  time by assuming that the objects radiate at 
Eddington luminosity $L_{Edd}$.  For a central BH of mass $M_H$, 
$L_{Edd}=4\pi G (M_H/M_\odot) c/\ka=1.3 \times 10^{38} 
(M_H/M_\odot)\ {\rm erg/sec},$ where $\ka$ is the Thomson opacity 
$\sim 0.4{\rm cm}^2 {\rm\ g}^{-1}$.  
As the AGN radiates at $\sim L_{Edd}$ we have
${\dot M}_{Edd}=2.2 \times 10^{-9} \ep^{-1}M_H {\rm yr}^{-1},$ 
where $\epsilon \lsim 1$ is the efficiency factor for radiation 
by accretion. Then 
$$
\tau_{vis}\ll \tau_{g}\lsim M_H/{\dot M_{Edd}}=4.5\epsilon\ {\times}10^{8}{\rm yr}.
\eqno(2)
$$

We must also compare $\tau_r$, the gravitational relaxation time of the 
IC region, with $\tau_o$, the  orbital time scale.  When the BH  
is small, its radius of influence $R_{In}\sim GM_{H}/\sigma^2$ satisfies 
$R_{In}<R_{IC}$, where $R_{IC}$ is the IC radius and $\sigma$ is the 
stellar velocity dispersion, so that $\tau_r$ is 
independent of $M_H$. When $R_{In}\ge R_{IC}$ the relaxation time depends on $M_{H}$. In the Fokker-Planck approximation, 
for the case $R_{In}<R_{IC}$ we have (Binney \& Tremaine 1987)
$$
\tau_r\sim 10^{11} {\rm yr}~~ (\sigma/100{\rm km}\
s^{-1})^3(10^5M_\odot\ {\rm pc}^{-3}/\rho_{_{IC}}),
\eqno(3)
$$
where $\rho_{_{IC}}$  is the IC mass density.
For $R_{In}\gg R_{IC}$ the relaxation time is
$$
\tau_r\sim \si^3/(G^2M_\odot \rho_{_{IC}})=\si^3(4\pi R_{In}^3/3)
/(G^2M_\odot^2 N) =4\pi G M_H^3/(3\si^3 M_\odot^2 N)
$$
$$
\sim10^{13}{\rm yr}~~(M_{H}/10^8M_\odot)^3(\si/100{\rm km\ s}^{-1})^{-3}
(N/10^{9})^{-1},
\eqno(4)
$$
where $N$ is the number of IC stars.

Because of $(2)$,(3) and (4), 
$\tau_o\ll \tau_H\ll \tau_r,$ where $\tau_H$ is the Hubble time 
$\sim 10^{10}$ yr and $\tau_o\lsim10^6\ {\rm yr}(R_{IC}/10{\rm pc})
(\sigma/ 100{\rm km\ s^{-1}})^{-1}$,  
the adiabatic approach is appropriate. 
Young (1980) considers the adiabatic evolution of an isothermal sphere
with a growing BH.  Quinlan et al. (1995) confirm the results
of Young (1980) and extend to non-isothermal spheres.  
As applied to the IC, these are reasonably consistent with a 
total mass dependence on radius given by
$$
M_{tot}(r)=M_H[1+m(r/R_{IC})^n],\  r\le R_{IC},
$$
$$
M_{tot}(r)=(M_H+M_{IC})r/R_{IC},\  r> R_{IC},
\eqno(5)
$$
where  $m\equiv M_{IC}/M_{H}$ is the ratio 
of BH to stellar core mass, and $n \sim 3$. 

\noindent
{\bf 3. Gas Density}

The surface gas density is given by $\Si(r)=2H(r)\rho(r)$, where
$H(r)$ is the half-scale height of the gas and $\rho(r)$ is the
gas density.  Following the standard treatment (e.g., Pringle 1981) 
we take the viscosity coefficient to be $\nu=\gamma v_T H,$
where we assume $\gamma\siml 1$ is the  viscosity parameter and 
$v_T$ is the turbulent eddy speed.  
Combining this with the conservation of gas mass and angular 
momentum, the surface density satisfies (Pringle 1981, Yi et al. 1994)
$$
d\Sigma/dr+f(r)\Sigma=g(r),
\eqno(6)
$$
where
$f(r)=(r^3\Omega'/\Om)'/(r^3\Om'/\Om)$, 
$g(r)=-[{\dot M}\Om/(2\pi \gamma v_T^2 r)]
[2(\Om/r\Om')+1]$, and $\Om$
is the angular velocity determined by the potential. 
The $'$ indicates $d/dr$. The solution of $(6)$ is given by
$
\Si(r)={\exp}\left[-\int^r_{r_0}ds f(s)\right]\left(\Si_0+\int_{r_0}^r ds 
g(s){\exp}\left[\int_{r_0}^s d\la
f(\la)\right]\right),$
where $\Si(r_0)=\Si_0$. To find the relationship between the 
surface density and the stellar mass density we note that 
for circular orbits, $\Om=(GM/r^3)^{1/2}$. 
Plugging in for $f(r)$ and $g(r)$, 
using $(5)$, $S\equiv r/R_{IC}$, and $\la \equiv m^{1/n}$G we have
$$
{\Si(r)\over \Si_0}={(1+mS^n)\over {R_{IC}^{2}([n-3]mS^{n+2}-3S^2)}}
\left[{{R_{IC}^{2}([n-3]mS_0^{n+2}-3S_0^2)}\over {1+mS_0^n}}
-{K\over \Si_0}\int_{m^{1/n}S_0}^{m^{1/n}S}d\la 
{{(n+1)\la^n+1}\over {(\la +\la^{n+1})^{1/2}}}\right],
\eqno(7)
$$
where $K={\dot M}m^{{1 \over 2}(1-1/n)}(G M_H R_{IC})^{1/2}
/(2\pi\gamma v_T^2)$.

The third term in $(7)$ is negative, so the second term gives the correct 
order of magnitude.  Fig. 1 shows the surface density at 
$r=R_{IC}\sim R_T$, the edge of torus, (outside of which the potential drops) 
for 4 values of $n$ as a function of $m$ and vice versa. 
For large $m$ and $n\sim 3$ the density can be 100 times the Keplerian 
(i.e. when $m\ll 1$) value. That 
$n\sim 3$ is consistent with Young (1980) and Quinlan et al. 
(1995).  This density enhancement can account for X-ray 
obscuration in Seyferts (Yi et al. 1994).

\noindent
{\bf 4. Rigid Rotation and Poloidal Field Growth}

As shown above, the enhanced density torus results 
from reduced differential rotation. We can calculate the 
reduction in $\Omega'$ by setting 
$\Om=(GM_{tot}(r)/r^3)^{1/2}$.  Using $(5)$ with $n=3$ gives
$\Om'(R_{IC})=-(3/2)G^{1/2}R_{IC}^{-5/2}M_H^{1/2}/(1+m)^{1/2}.$
For $m>50$ this gives a factor $> 7$ reduction in $\Omega'$
 from the Keplerian value.  This region may be important for 
the dynamo generation of PMF for radio sources as we now describe.

The mean field dynamo theory (Moffatt 1978, Parker 1979)
splits the velocity and the magnetic field into mean $({\bbfv},{\bbfb})$
and fluctuating $({\bf v}',{\bf b}')$ components.
The time evolution of 
mean magnetic field is given by (Moffatt 1978, Parker 1979)
$$
\partial{\bfbb}/\partial t=
\nabla \times [{\bbfv}\times {\bfbb}+\a {\bfbb}-(\lambda_M+\beta)\nabla 
\times{\bfbb}],
\eqno(8)
$$
where $\a$ and $\beta$ are the helicity and diffusion dynamo coefficients 
and are functions of the turbulent velocity. 
A non-vanishing $\a$ is the result of buoyant eddies rising 
in an upwardly decreasing density gradient, while conserving their angular 
momenta.  The magnetic viscosity, $\lambda_M$, satisfies 
$\lambda_M\ll \beta$.  

For small $\Om'$, the ``$\a^2$ dynamo'' 
(Moffatt 1978) is favored, because the maximum growth rate depends on 
$\a^2$ as we will see. Sufficient reduction of $\Om'$ 
means that the radial PMF produced by the $\alpha$ 
effect is not sheared into toroidal field. Simulations 
(e.g., Donner \& Brandenburg 1990) show that dipole modes are 
favored in such a dynamo, in contrast to the $\a-\Om$ dynamo often 
applied to disks (Parker 1979).  

To estimate when the $\alpha^2$ dynamo is favored,
we work in cylindrical coordinates $(r,\phi,z)$ and write
${\bfbv}=r\Om(r,z){\hat{\bf e}}_\phi,\ {\bfbb}={\bar B}_\phi(r,z)
{\hat{\bf e}}_\phi+{\bfbb}_P,$ where $P$ indicates the poloidal 
$(r,z)$ component, and $\phi$ indicates the toroidal component.
${\bfbb}_P=\nabla\times {\bar A}(r,z){\hat{\bf e}}_\phi$.  
Assuming $\a$ and $\beta$ are constant, $(8)$ can be written  
(Moffatt 1978)
$$
\partial {\bb}_\phi/\partial t=r({\bfbb}_P {\cdot} \nabla)\Om-
\a (\nabla^2-r^{-2}){\ba}+\beta(\nabla^2-r^{-2})\bb_\phi,
\eqno(9)
$$
and
$$
\partial \ba/\partial t=\a \bb_\phi+
\beta(\nabla^2-r^{-2})\ba.
\eqno(10)
$$ 
An $\a^2$ dynamo will dominate the  $\a-\Om$ dynamo when the second source
term on the right of $(12)$ dominates the first, that is
when $ \a/r\gg |r\Om'|.$  From Parker (1979), we have $\a \sim 0.4v_T$.  
For a turbulently supported dust torus, observations require 
$v_T/V_\phi\sim 0.5$ where $V_\phi=r\Om$ (Krolik \& Begelman 1988, 
Urry \& Padovani 1995).  Using these and $(5)$, 
the requirement near $r=R_{IC}\sim R_{T}$ becomes
$0.20M_H^{1/2}(1+m)^{1/2}>(3/2)M_H^{1/2}(1+m)^{-1/2}$,
or 
$$
m\gg6.5.
\eqno(11)
$$
When $(11)$ is satisfied, we can ignore the first term on the
right of $(8)$ near $r=R_{IC}$. We capture the essence of an 
$\a^2$ dynamo, by considering the case when the $z$-variation dominates
and assuming solutions of the form 
$\bb_\phi,\ba\ \propto r{\exp}(\ga t+k_zz)$.
Plugging these into $(9)$ and $(10)$ gives
$$ 
\ga \bb_\phi= \a \ba k_z^2-\be\bb_\phi k_z^2
\eqno(12)
$$
and
$$ 
\ga {\ba}=\a{\bb}_\phi-\be\ba k_z^2,
\eqno(13)
$$
so that the growth rate is given by
$$
{\rm Re}[\ga]=-\be k_z^2/2+3k_z\a/2.
\eqno(14)
$$
The growth rate is positive if $k_z<3\a/\be$.  
The maximum growth rate is ${\rm Re}[\ga]_{max}=(9/8)\a^2/\beta$, showing 
the $\a^2$ dependence. The second term on the right of $(14)$ provides 
a more conservative estimate.  

Let us see why an\ $\a^2$ dynamo favors PMF.  
For an AGN torus, the fluctuation scale is determined by the size of dust 
containing clouds. The dust must be in clouds because
it could not survive if the random velocities of $\ge 100$km/s were 
thermal. Thus the cloud size $r_c$, satisfies $r_c<R_T$, where the torus 
radius $R_T$ is the scale for variation of the mean quantities.
We can estimate the dust cloud size from observations of the CNR 
of our Galaxy (Genzel \& Townes 1987), which is thought to be 
similar to the AGN dust tori (Krolik \& Begelman 1988). These observations 
(Genzel \& Townes 1987) show clouds with $0.1 \lsim r_c \lsim 0.25$pc.  
Now we estimate the PMF produced: Setting $Re[\ga] \sim k_z\a$ 
and using $k_z\ba\sim {\bb}_P$, 
with $\a \sim 0.4 v_T$ and $\be\sim (1/3) r_cv_T$ in $(12)$ and $(13)$
gives ${\bb}_P\sim \bb_\phi$ for the $\a^2$ dynamo.  The analogous equations
to $(12)$ and $(13)$ for the $\a-\Om$ dynamo, derived by keeping the first 
term on the right of $(8)$ and dropping the term linear in $\a$, 
give $[{\bb}_P/\bb_\phi]_{\a-\Om}\sim r_c/R_T$.  Thus
$[B_p/\bb_\phi]_{\a^2}/[B_p/\bb_\phi]_{\a-\Om}\sim R_T/r_{c} \gsim\ 50-100,$
showing that the $\a^2$ dynamo favors PMF in comparison
to the $\a-\Omega$ dynamo.  This ratio is important for jet models, 
particularly when the resulting luminosity depends on $B_p^2$ 
(e.g., Blandford \& Znajek 1977). Thus a factor of 50-100 in the field 
corresponds to a factor of $2.5\ts 10^3$ to $10^4$ in the jet luminosity.

\noindent
{\bf 5. Poloidal Field in AGN}

For PMF to be produced by a working torus dynamo, 
$(11)$ must hold. 
In addition, the dynamo growth 
time must be shorter than the torus lifetime. That is,
$$
\tau_d<\tau_{g}M_T/M_H.
\eqno(15)
$$
Note $\tau_d$ for the torus must be less than the 
field diffusion time: 
$\tau_{d}< (10 {\rm pc})^2/\beta\sim 100
{\rm kpc}^2/(100{\rm km s^{-1}\ }10{\rm kpc})\sim 10^5{\rm yr}$.
For a density $\rho_T\sim 5\ts 10^{-18}$g/cm$^3$ (Krolik \& Begelman 1988)
and radius $5$pc with height 2.5pc, $M_T\sim  10^7 M_{\sun}$.
$~$From $(2)$ with $\ep\sim 0.1$, $\tau_{g}\sim 5\ts 10^7$yr. Thus
violating $(15)$ requires the  extreme case of 
$M_H>10^4 M_T$ so that $(8)$ is more stringent.  

The larger $m$ is the greater the density enhancement and produced PMF.  
Equipartition between turbulent and magnetic energy densities gives
an upper limit to the field. 
For $\rho_T\sim 5 \ts 10^{-18}{\rm g/cm^3}$ (Krolik \& Begelman 1988), 
corresponding to $m\gsim 50$, and $v_T\sim 100$km/s the turbulent energy 
is $(1/2)\rho_T v_T^2\sim 2.4\ts10^{-4} {\rm erg/cm^3}$. Setting this 
equal to  $B_P^2/8\pi$ we have $B_P\siml 8\ts 10^{-2}{\rm G}$. 
The field is accreted to the central engine as the torus depletes.
The radial component of PMF is then sheared, but only that 
fraction of the field at a much lower scale height than
that of the $\sim 5-10$pc torus.  The z-component is unaffected by the shear.  
An estimate of the 
accreted PMF can then be made from flux freezing.  For an 
ion-electron accretion disk with height to radius ratio $H_d/R_d \sim 1/50$ 
and density $\rho_{disk}\sim 10^{-8}$ cm$^{-3}$ at $r\sim 10^{14} cm$
(e.g., Celotti et al. 1992), flux conservation implies that PMF 
will accrete to 
$B_{P, disk}< B_{P,T}[\rho_{disk}(H_d/R_d)/\rho_T (H_T/R_T)]^{2/3}
\sim 8\ts 10^{-2}[10^{-8}(1/50)/5 \ts 10^{-18}(1/2)]^{2/3}\sim 1.5\ts 10^4$G, 
in agreement with standard estimates (Begelman et al. 1984).  

Only when the torus depletes by accretion
can the field produced there move to the central engine and 
play its role in jets.
As any jet formation time in the engine is likely much shorter than
the torus depletion time, only during the latter time scale  
can a jetted object show both a jet and a torus.
For a $10^6M-10^7M\sun$ hole accreting at the Eddington rate 
with $\ep\sim 0.1$ and a torus mass of $10^7M\sun$, this accretion period
lasts $\siml 10^7-10^8$ yr. Note also that a
torus scale dynamo need not determine the final scale of the magnetic 
field, but would mediate the initial energy extraction from the rotating 
central source.  The initial jet flow could drag the resulting field to 
kpc-Mpc scales as in Blandford \& Rees (1974). 

\noindent
{\bf 6. Conclusions}
 
Reduced $\Om'$ tori in the central $\lsim 10$ pc
regions of AGN can obscure X-rays and incur dynamo production 
of PMF likely required for jets, when $m \gsim 50$ initially. 
PMF growth in the torus allows angular momentum transport, and the torus 
will accrete to the central engine carrying its field. 
PMF transport and torus depletion are associated processes and  
jetted objects would be less likely to show obscuring tori. 
Jet-free AGN would either have an intial $m<50$ or have their jets beamed
away from us.  There are a few radio loud quasars (RLQ) or Seyferts with
inferred BH masses $>10^8M\sun$ (e.g. NGC 5548, Krolik et. al. 1991).
The absence of direct evidence for dust tori in the latter may be consistent
with our paradigm, but may require non-adiabatic analysis.

\ni{\bf Acknowledgements}:  We thank George Field for useful discussions 
during the early stage of this work. I.Y. acknowledges the support of SUAM 
Foundation.  

\ni Antonucci, R. 1993, ARA\&A., {\bf 31}, 473

\ni Appl, S. \& Camenzind,M. 1992 , A\&A, {\bf 270}, 71

\ni Begelman, M.C., Blandford, R.D., \& Rees, M.J, 1984, Rev. Mod. Phys., 
{\bf 56}, 256

\ni Blandford, R.D. \& Znajek, R.P. 1977, MNRAS, {\bf 179}, 433

\ni Blandford, R.D. \& Rees, M.J. 1974, MNRAS, {\bf 169}, 395

\ni Celotti, A., Fabian, A.C.,  \& Rees, M.J., 1992, M.N.R.A.S, {\bf 255}, 
419

\ni Donner, K.J., \& Brandenburg, A., 1990,  A\&A {\bf 240}, 289

\ni Duschl, W. J. 1989, MNRAS, {\bf 240}, 219

\ni Genzel, R. \& Townes, C.H., 1987, ARA\&A, {\bf 25}, 377

\ni Krolik, J.H. \& Begelman, M.C. 1988, ApJ, {\bf 329}, 702

\ni Krolik, J.H., Horne,K., Kallman,T.R., Malkan,M.A., Edelson,R.A. \&
Kriss,G.A., 1991, Ap.J. {\bf 371}, 541

\ni Lovelace, R.V.E., Wang, J.C.L., \& Sulkanen, M.E. 1987, ApJ, {\bf 315},
504

\ni Lynden-Bell, D. 1995, Cambridge preprint CAP-9510001 

\ni Moffat, H.K. 1978, {\sl Magnetic Field Generation in Electrically 
Conducting Fluids}, (Cambridge: Cambridge Univ. Press)

\ni Oort, J.H. 1978, ARA\&A, {\bf 15}, 295

\ni Parker, E.N. 1979, {\sl Cosmical Magnetic Fields}, (Oxford: 
Clarendon Press)

\ni Pringle, J.E. 1981, ARA\&A, {\bf 19}, 137

\ni Quinlan, G.D., Hernquist, L.,\& Sigurdsson, S. 1995, ApJ, {\bf 440}, 554

\ni Rees, M.J. 1984, ARA\&A, {\bf 22}, 471

\ni Urry, C.M., \& Padovani, P., 1995, PASP, {\bf 107}, 803

\ni Yi, I., Field, G.B., \& Blackman, E.G. 1994, ApJ, {\bf 432}, L31

\ni Young, P. 1980, ApJ, {\bf 242}, 1232

\bigskip
\centerline{\bf Figure Caption}
\bigskip

\noindent
Figure 1: (a) Surface density at the torus' outer edge as function of $m$ 
for $n=0.01,\ 1,\ 2.7\ ,3$ going from the bottom to the top curves. 
(b) Surface density as a function of $n$ for $m=0.01,\ 1,\ 10,\ 100$ 
from the bottom to top curves.

\end